# Quantum effects in piezoelectric semiconductor plasmas : Solitons and transmission feasibility

**Abhishek Yadav and Punit Kumar[*]**

*Department of Physics, University of Lucknow, Lucknow-226007, India*
*Email: kumar_punit@lkouniv.ac.in (Corresponding Author)*

## ABSTRACT

A study of coupling of the lattice ion vibrations, with the electron waves in a piezoelectric semiconductor quantum plasma is presented. The nonlinearities have been studied and the solitons have been analysed. The theory has been built using the quantum hydrodynamic (QHD) model, incorporating the effects of Fermi pressure, quantum Bohm and exchange correlation potentials. The dispersion relation for the coupling has been set up. A set of nonlinear evolution equations has been established using the two-time scale theory and soliton solution for coupled nonlinear evolution equations has been obtained using the modified quantum Zakharov equations. The solitons are found to be have a cusp profile. It is also found that soliton's field amplitude increases significantly with particle density and coupling strength in piezoelectric semiconductor quantum plasma.

**Keywords:** Piezoelectric semiconductor, Quantum plasma, QHD model, Solitons.

## 1. Introduction

The piezoelectric effect was initially observed in various solid materials [1]. Later, based on thermodynamic principles, the reverse piezoelectric effect, which involves changing the size of a crystal through the application of an electric field, was proposed [2-4]. A significant contribution to the field has been made by formulating the theory of elasticity [5] and demonstrated that the electrical and mechanical states of a crystal could be described by piezoelectric constants. An advancement to the piezoelectric phenomenon in semiconductors was made by developing a linear theory encompassing both intrinsic and extrinsic semiconductors, accounting for the effects of carrier diffusion, trapping, and drift [6]. Since then, piezoelectric semiconductors have garnered extensive industrial applications and in micro-electro-mechanical systems (MEMS). The utilization of piezoelectric materials for energy harvesting has been a subject of investigation [7-9]. With the emergence of nanotechnology, piezoelectric materials have ventured into new domains, spurring active exploration in pursuit of novel applications [10-12].

Within semiconductors, electron and holes obey Fermi-Dirac statistics rather than the classical Boltzmann distribution. This distinction becomes crucial in the miniaturization of electronic components where the use of semiconductors relies on the precise adjustment of charge carrier's De Broglie wavelengths to match the interparticle spacing in doping profiles. Consequently, typical quantum mechanical effects are expected to exert a fundamental influence on the behaviour of forthcoming electronic components. For modern physicists dealing with quantum structures such as quantum wells, quantum wires, and quantum dots, understanding both the linear and nonlinear characteristics of waves and instabilities driven by carrier dynamics in semiconductors is essential. Additionally, the development of wireless electronic devices has raised significant concerns regarding wireless power sources. In this context, piezoelectric semiconductors emerge as natural candidates for the conversion of mechanical stress into electric field and vice versa. This conversion is facilitated by the coupling between lattice ion vibrations and electrokinetic modes through piezoelectricity [13,14], which has sparked substantial and sustained interest.

The quantum effects can be explored through quantum hydrodynamic (QHD) equations, with the Bohm potential encapsulating the quantum tunneling along with other quantum effects [15-17]. The QHD model has successfully derived the dielectric tensor and dispersion relations for both the longitudinal and transverse electromagnetic waves in semiconductor quantum plasmas [18]. Notably, quantum corrections have a significant impact on longitudinal waves, leading to rapid decay due to Landau damping [19]. Moreover, quantum surface modes can manifest at the interface between plasmas and vacuum in magnetized electron-hole semiconductor plasmas, as revealed by the QHD model [20]. This is particularly significant, as the quantum effects lower the threshold electric field for parametric amplification making it easier to achieve the necessary pump electric field, especially in unmagnetized piezoelectric semiconductors [21]. More recent investigations have explored the two-stream instability in quantum semiconductor plasmas using the QHD model [22]. The quantum effects reduce the Debye length giving rise to a quasi-quantum lattice of colloid ions on quantum scales [23]. Additionally, instability in electron beam-pumped GaAs semiconductors can be attributed to the excitation of electron-hole pairs [24]. Various other aspects have extensively explored acoustic wave behaviour within this regime. For instance, a comprehensive investigation has been done into the propagation of longitudinal acousto-electric waves in colloids embedded in semiconductor plasmas, utilizing the quantum hydrodynamic (QHD) model [25]. The linear and nonlinear properties of ion-acoustic waves in a three-component quantum plasma using the QHD equations have been examined [26]. The modulation of quantum positron acoustic waves has been studied in detail [27].

The present work investigates the coupling between lattice ion vibrations and electron waves in a piezoelectric semiconductor quantum medium, aiming to analytically study solitary structures for the coupled nonlinear evolution equation using the Zakharov approach. This study employs the quantum hydrodynamic model (QHD) model, which effectively examines the behavior of electrons and lattice ions in such systems, although its applicability is limited to scales significantly larger than the Fermi length of the species involved. The QHD model boosts several advantages over kinetic models, including numerical efficiency and the ability to directly incorporate macroscopic variables like momentum and energy, facilitating the study of nonlinear phenomena in quantum plasmas [28-32]. As semiconductor devices shrink in sizes, quantum effects become crucial for accurately modeling charge carriers. The successful application of QHD model in semiconductor plasma studies has demonstrated its capability to predict behaviours unattainable by classical models, including the dynamics of electron-acoustic waves and surface plasmon oscillations [33,34]. Moreover, the QHD model is adapt at modeling short scale collective phenomena, such as instabilities and nonlinear interactions in dense plasmas prevalent in semiconductors [29,35].

In Section 2, the fundamental theoretical formulation required for the study of the coupled lattice-electron mode in piezoelectric semiconductor quantum plasma has been build up. This encompasses the relevant equations and models employed to depict the system. A quantum dispersion relation has been derived for the coupled lattice-electron mode, thereby offering insights into the behavior of these modes in the linear regime. Section - 3 delves into the nonlinear analysis and soliton solution. A set of nonlinear evolution equations that describe the coupling between the lattice ion waves and electron waves in the piezoelectric semiconductor quantum plasma are derived by utilizing two-time scale theory. Finally, in Section 4, summary along with discussion is presented.

2. **Ion-electron modes**

We consider the following piezoelectric equations of state [36],

$$\vec{T} = \tau \vec{S} - \beta \vec{E}, \tag{1}$$

$$\vec{D} = \varepsilon \vec{E} + \beta \vec{S}. \tag{2}$$

The QHD fluid equations [37],

$$\frac{\partial \vec{v}_\alpha}{\partial t} + \vec{v}_\alpha \cdot \nabla \vec{v}_\alpha = -\frac{\nabla P_{\alpha F}}{m_\alpha n_\alpha} - \frac{e_\alpha}{m_\alpha} \vec{E} + \frac{\hbar^2}{2m_\alpha^2} \nabla \left( \frac{1}{\sqrt{n_\alpha}} \nabla^2 \sqrt{n_\alpha} \right) + \frac{\nabla V_{xc}}{m_e}, \tag{3}$$

$$\frac{\partial n_\alpha}{\partial t} + \nabla \cdot \left( n_\alpha \vec{v}_\alpha \right) = 0. \tag{4}$$

In equations (1) and (2), $T, \tau, S, \beta, E$ and $D$ represent stress, elastic constant, strain, piezoelectric coupling parameter, electric field and displacement respectively. Eq. (1) signifies that an applied electric field causes a strain in the plasma medium, effectively driving mechanical stress which induces a change in electric displacement associated with generation of piezoelectric current [38-40]. Eqs. (3) and (4) correspond to momentum and continuity equations for plasma species $\alpha$, where $n_\alpha$ is the particle density, $e_\alpha$ is the charge on the particle and $m_\alpha$ is the particle's effective mass. The second term, on the left hand side of eq. (3) is the convective derivative of the velocity. The first term on RHS of eq. (3) is the force due to Fermi pressure $P_{\alpha F} = \frac{m_e v_F^2}{3n_{0\alpha}^2} n_\alpha^3$, where $v_F = \sqrt{\frac{2k_B T_F}{m_e}}$, is Fermi velocity. The second term is the force under the influence of an applied electric field, the third term corresponds to the force of the quantum Bohm potential arising from quantum corrections in density fluctuations and effects the phase and group velocities in semiconductor plasma [41,42]. It is crucial for stability analysis and plays a significant role in wave amplification [43]. The fourth term describes the influence of exchange correlation, where $V_{xc}$ is the exchange and correlation potential [44],

$$V_{xc} = -\frac{0.985 e_\alpha^2}{\varepsilon}(n_\alpha)^{1/3}\left[1+\frac{0.034}{a_B(n_\alpha)^{1/3}}\ln\left(1+18.37 a_B(n_\alpha)^{1/3}\right)\right], \quad (5)$$

where, $\varepsilon(=\varepsilon_0 \varepsilon_L)$ is the permittivity of the medium with lattice dielectric constant $\varepsilon_L$, the parameter $a_B$ is the Bohr's atomic radius and $n_\alpha$ is the particle density. It is noteworthy that all the variables in eqs. (1) and (2) are scalar quantities, as we are considering the propagation in the $x$-direction only. The strain $S$ is written as $S = \partial u/\partial x$, where $u$ is the physical displacement of particles. The equation (1), gives us the wave equation in elastic medium,

$$\frac{\partial^2 \vec{u}}{\partial t^2} = c_s^2 \frac{\partial^2 \vec{u}}{\partial x^2} - \frac{\beta}{\rho}\nabla.\vec{E}. \quad (6)$$

where, $c_s = \sqrt{\tau/\rho}$ is the speed of sound and $\rho$ is the density of medium.

**2.1 The electron dynamics**

We perturb eqs. (3) and (4) in orders of the piezoelectrically generated electric field, where all the varrying parameters take the form

$$f = f_0 + f^{(1)}$$

with $f_0$ representing the unperturbed value and $f^{(1)}$ is the perturbation term. The first order momentum and continuity equations for an electron can now be written as,

$$\frac{\partial \vec{v}_e^{(1)}}{\partial t} = -\frac{\nabla P_{eF}^{(1)}}{m_e n_{0e}} - \frac{e}{m_e}\vec{E}^{(1)} + \frac{\hbar^2}{4m_e^2 n_{0e}}\nabla\left(\nabla^2 n_e^{(1)}\right) + \frac{A_{xc}\nabla n_e^{(1)}}{m_e n_{0e}}, \quad (7)$$

and

$$\frac{\partial n_e^{(1)}}{\partial t} + n_{0e}\nabla.\left(\vec{v}_e^{(1)}\right) = 0. \quad (8)$$

where, $A_{xc} = \dfrac{0.328 e^2 n_{0e}^{1/3}}{m_e \varepsilon}\left(1 + \dfrac{0.625}{1 + 18.37 a_B n_{0e}^{1/3}}\right)$ is the term corresponding to exchange and correlation potential.

Assuming all the perturbed quantities to vary as $e^{i(kx-\omega t)}$, we get the first order perturbed velocity $v_{ex}^{(1)}$ as,

$$v_{ex}^{(1)} = \dfrac{\left(\dfrac{e}{m_e}\right)}{i\left(\omega - v_F^2\left(1 + \dfrac{k^2 v_F^2}{4\omega_p^2}H^2\right)\dfrac{k^2}{\omega} + \dfrac{k^2}{\omega m_e}A_{xc}\right)} E_x^{(1)}, \qquad (9)$$

and the perturbed electron density $n_{ex}^{(1)}$ as,

$$n_{ex}^{(1)} = \dfrac{n_{0e} k}{\omega} v_{ex}^{(1)}. \qquad (10)$$

Using Maxwell's relation $\nabla \cdot D_{ex}^{(1)} = -e n_{ex}^{(1)}$, we get the perturbed electric displacement vector as,

$$D_x^{(1)} = \dfrac{\left(\dfrac{n_{0e} e^2}{m_e \omega}\right)}{\left(\omega - v_F^2\left(1 + \dfrac{k^2 v_F^2}{4\omega_p^2}H^2\right)\dfrac{k^2}{\omega} + \dfrac{k^2}{\omega m_e}A_{xc}\right)} E_x^{(1)}. \qquad (11)$$

where, $\omega_p \left(= \sqrt{\dfrac{n_{0e} e^2}{m_e \varepsilon}}\right)$ is the electron plasma frequency and $H\left(= \dfrac{\hbar \omega_p}{2 k_B T_F}\right)$ is a non-dimensional quantum parameter. Eq. (11) represents the desired perturbed electric displacement component for free electron of plasma medium in terms of perturbed electric field component $E_x^{(1)}$. It accounts for the response of electron in the plasma medium on application of electric field with the inclusion of quantum effects such as electron degeneracy in description of electron dynamics in the quantum plasma medium.

## 2.2 The ion dynamics

In the similar manner as done in the previous section 2.1, the dynamical eqs. for ion can now be written as,

$$\frac{\partial \vec{v}_i^{(1)}}{\partial t} = \frac{e}{m_i}\vec{E}^{(1)}, \tag{12}$$

$$\frac{\partial n_i^{(1)}}{\partial t} + n_{0i}\nabla\cdot\left(\vec{v}_i^{(1)}\right) = 0, \tag{13}$$

and

$$\frac{\partial^2 \vec{u}_i^{(1)}}{\partial t^2} = c_s^2 \frac{\partial^2 \vec{u}_i^{(1)}}{\partial x^2} - \frac{\beta}{\rho_i}\nabla\cdot\vec{E}^{(1)}. \tag{14}$$

The neglect of some quantum terms in ion dynamics within the context of quantum plasma is based on considerations that ions are much more massive than electrons i.e., $m_e/m_i = 1/1836$. Due to this large mass, ions are treated classically and electron are subjected to quantum effect and the thermal de-Broglie wavelength of ions is typically much smaller than electrons which allows for the classical treatment of ion dynamics [34]. Also the thermal de-Broglie wavelength of ions is not comparable to Debye length of plasma, justifying their classical treatment [45]. From the above equations, equation of motion for lattice ion comes out to be,

$$\left(\omega^2 - c_s^2 k^2\right)u_{ix} = \frac{\beta}{\rho_i}ikE_x^{(1)}. \tag{15}$$

In the absence of piezoelectricity $(\beta = 0)$, the above equation describes the usual wave equation. However, its presence is essential for the coupling of lattice ion vibrations with the electron waves through the electric field $\vec{E}^1$. The modified Poisson's equation for the charge separation is,

$$\nabla\cdot\vec{E} = -\frac{e}{\varepsilon}n_e - \frac{\beta}{\varepsilon}\frac{\partial^2 \vec{u}_i}{\partial x^2}. \tag{16}$$

The displacement vector $D_x^{(1)}$ of lattice ion can be obtained by mutually solving eqs (2), (15) and (16),

$$D_x^{(1)} = \left(\frac{\omega^2 - c_s^2 k^2 - \dfrac{\beta^2 k^2}{\rho_i \varepsilon}}{\omega^2 - c_s^2 k^2}\right)\varepsilon E_x^{(1)}, \tag{17}$$

where, $\varepsilon(=\varepsilon_0\varepsilon_L)$ is the permittivity of the medium with $\varepsilon_0$ and $\varepsilon_L$ being the permittivity in free space and lattice dielectric constant respectively. Eq. (17) represents the desired perturbed electric displacement component for lattice ions in terms of perturbed electric field component $E_x^{(1)}$. It accounts for the response of ions in the plasma medium on the application of electric field.

### 2.3 Coupling of electron - ion modes

We now proceed to study the coupling of electron-ion modes due to the piezoelectric field in n-type piezoelectric semiconductor quantum plasma. Using eqs. (11) and (17), we obtain the coupled dispersion relation in terms of quantum parameter $H$,

$$\left(1-\frac{\omega_p^2}{\left(\omega^2-v_F^2\left(1+\frac{k^2 v_F^2}{4\omega_p^2}H^2\right)k^2+\frac{k^2}{m_e}A_{xc}\right)}\right)\left(\omega^2-c_s^2 k^2\right)=\frac{\beta^2 k^2}{\rho_i \varepsilon}. \tag{18}$$

In the above equation, first term on the left-hand side represents the electron plasma mode, while the second term represents the lattice acoustic mode in piezoelectric semiconductor quantum plasma. The term on the right-hand side is the coupling term that accounts for the interaction between the electron plasma and lattice acoustic modes. It is interesting to note that in the absence of piezoelectricity $(\beta=0)$, the coupling parameter vanishes and the dispersion relation decouples into two independent modes, the lattice acoustic and the Langmuir modes. The electron plasma and lattice vibrations evolve independently of each other, without any interaction, or coupling between them

$$\left(\omega^2-\omega_p^2-v_F^2\left(1+\frac{k^2 v_F^2}{4\omega_p^2}H^2\right)k^2+\frac{k^2}{m_e}A_{xc}\right)=0. \tag{19}$$

$$\left(\omega^2-c_s^2 k^2\right)=0. \tag{20}$$

In the numerical analysis to follow, the parameter chosen are for an n-type InSb semiconductor; $\varepsilon_0=8.85\times10^{-12}\,C^2/N.m^2$, $\varepsilon_L=17.54$, $\varepsilon=1.55\times10^{-10}$, $n_{e0}=10^{28}/m^3$, $\rho_i=5.8\times10^3$, $m_e=0.014 m_0$, $m_0=9.1\times10^{-31}\,Kg$, $c_s=2500\,m/s$, $T=77K$ [46-50] and the

piezoelectric coupling constant $\beta$ for such type of materials ranges from $0.045\,C/m^2$ to $0.35\,C/m^2$ [21,46,51].

Fig. 1 shows the variation of normalised wave frequency $\omega/\omega_p$ with respect to normalized propagation vector $kc/\omega_p$. The solid line shows the variation in quantum plasma, while the dashed line shows the trend in absence of quantum effects $(\hbar \to 0)$. It is evident from the figure that the wave frequency is reduced by about 9% in quantum plasma as compared to the case where quantum effects are absent. It is due to the dominance of Fermi pressure over thermal pressure. This dominance of Fermi pressure results in a higher number of energy levels, which introduces degeneracy, which in turn causes reduction in wave frequency.

In Fig 2. the variation of normalised wave frequency $\omega/\omega_p$ with normalized propagation vector $kc/\omega_p$ is shown for different values of quantum parameter $H$. The solid, dashed and dotted line shows the variation for $H=0.066$, $H=0.045$ and $H=0.030$ respectively. The wave frequency is increases by 17% for $H=0.045$ in comparison to $H=0.030$ and by 13% for $H=0.066$ in comparison to $H=0.045$. This is due to the concurrent influence of Fermi pressure and quantum Bohm potential, as quantum parameter $H$ shows their combined effect. Thus, we conclude that the transmission of power increases for the same wave transmission due to quantum effects involving quantum phenomena of tunneling highlighting the impact of quantum mechanics.

### 3. Soliton evolution

We now proceed to explore the nonlinear evolution for the coupling between electron plasma waves and lattice ion vibrations. These two modes exhibit excitation at different time scales due to the difference in mass between electrons and ions. Considering the presence of these two-time scales, we are able to derive a nonlinear evolution equation that captures the dynamics of the coupled modes. This nonlinear evolution equation allows us to study the interactions and behavior of the system beyond the linear approximation [52, 53],

$$n_i(x,t) = n_0 + n_{si}(x,t), \tag{21}$$

$$n_e(x,t) = n_0 + n_{se}(x,t) + n_{fe}(x,t), \tag{22}$$

$$v_i(x,t) = v_{si}(x,t), \tag{23}$$

$$v_e(x,t) = v_{se}(x,t) + v_{fe}(x,t), \tag{24}$$

$$u_i(x,t) = u_{si}(x,t), \tag{25}$$

$$E(x,t) = E_s(x,t) + E_f(x,t), \tag{26}$$

where, $s$ and $f$ represents slow and fast parts respectively, while $n_0$ is time independent equilibrium particle density which is same for both electrons and ions as per quasineutrality condition. In degenerate electrons, the fast time scale corresponds to the rapid oscillations of the degenerate electron gas, where electron dynamics are dominated by collective interactions and quantum effects such as Fermi pressure and the Bohm potential. These fast oscillations, typically at the plasma frequency, are due to the high mobility of electrons, allowing them to respond quickly to electric fields and disturbances. The slow time scale arises from nonlinear interactions and ponderomotive forces, where the collective electron density variations evolve more gradually, influenced by the slow changes in the electric field or the overall system's energy distribution. The two time scales allow the model to capture both the immediate response of the electron gas and its long-term evolution under nonlinear effects [52, 53].

The Poisson's equation, in two-time scales is

$$\nabla.E = -\frac{e(n_{ef})}{\varepsilon} - \frac{e(n_{se} - n_{si})}{\varepsilon}. \tag{27}$$

The first term on R.H.S. in the above equation represents the fast component of the field, which corresponds to the rapid oscillations in electron number density. The second term represents the slow component of the field, accounting for the slow time scale non-neutrality. It is important to note that while the motion of lattice ions is only slowly varying due to their large mass, the electron motion can adapt to both the slow and fast time scales. With the help of eqs. (3) and (21) – (27), we get the following expression for the high-frequency part,

$$\frac{\partial^2 E_f}{\partial t^2} + \omega_p^2.E_f - v_F^2\nabla(\nabla.E_f) + \frac{\hbar^2}{4m_e^2}\nabla^3(\nabla.E_f) + \frac{A_{xc}}{m_e}\nabla(\nabla.E_f) + c^2(\nabla\times(\nabla\times E_f)) = \omega_p^2\left(\frac{n_{se}}{n_0}\right)E_f. \tag{28}$$

The above equation is a nonlinear equation and the nonlinearity arises through the interaction between the slowly varying electron number density $n_{se}$, and the rapidly oscillating field $E_f$. The fast-time scale electric field is assumed to vary as [52, 53]

$$E_f = \frac{1}{2}\left[\tilde{E}(x,t)e^{-i\omega_0 t} + c.c.\right], \tag{29}$$

where c.c. refers to complex conjugate. $\tilde{E}(x,t)$ is the slowly varying complex amplitude of rapidly oscillating field. Assuming $\omega_0 \approx \omega_p$ for the long wavelengths and small amplitude electron wave, we approximate $\omega_0^2 - \omega_p^2 = 2\omega_p \Delta\omega$ and $\Delta\omega = \omega_0 - \omega_p$. The second order time derivative of eq. (29) gives,

$$\frac{\partial^2 E_f}{\partial t^2} = \frac{1}{2}e^{-i\omega_0 t}\left(-\omega_0^2 \tilde{E} - 2i\omega_0 \frac{\partial \tilde{E}}{\partial t} + \frac{\partial^2 \tilde{E}}{\partial t^2}\right). \tag{30}$$

Since, we have assumed $\tilde{E}(x,t)$ to be slowly varying amplitude, we can neglect its second order time derivative in eq. (28) to obtain the complex nonlinear evolution equation,

$$i\frac{\partial \tilde{E}}{\partial t} + \left(\frac{v_F^2}{2\omega_0} - \frac{c^2}{2\omega_0}\right)\nabla(\nabla.\tilde{E}) - \frac{\hbar^2}{8\omega_0 m_e^2}\nabla^3(\nabla.\tilde{E}) - \frac{A_{xc}}{2\omega_0 m_e}\nabla(\nabla.\tilde{E}) + \frac{c^2}{2\omega_0}\nabla^2 \tilde{E}$$
$$+ \left(\nabla\omega - \frac{\omega_p . n_{se}}{2n_0}\right)\tilde{E} = 0. \tag{31}$$

The above equation represents the slow variation of the local electron number density in relation to the complex amplitude $\tilde{E}(x,t)$ of the rapidly oscillating electric field. We proceed to consider the slow frequency component of the electrons by averaging over the fast oscillations which gives,

$$\frac{\partial v_{se}}{\partial t} + \frac{e}{m_e}E_s + \frac{v_F^2}{n_0}\nabla n_{se} - \frac{\hbar^2}{4m_e^2 n_0}\nabla^3 n_{se} + \frac{e^2}{4m_e^2 \omega_p^2}\nabla|\tilde{E}|^2 - \frac{A_{xc}\nabla n_{se}}{m_e n_0} = 0, \tag{32}$$

$$\nabla.E_s = -\frac{e}{\varepsilon}n_{se} - \frac{\beta}{\varepsilon}\frac{\partial^2 u_i}{\partial x^2}, \tag{33}$$

$$\frac{\partial n_{se}}{\partial t} + n_0 \nabla.v_{se} = 0. \tag{34}$$

In eq. (32), the nonlinear term proportional to $\nabla |\tilde{E}|^2$ arises from the convective derivative term, as described by Thornhill and Ter Haar [54]. By eliminating $E_s$ and $v_{se}$ from equations (29)-(31), we get

$$\frac{\partial^2 n_{se}}{\partial t^2} + \omega_p^2 . n_{se} + \frac{\beta \omega_p^2}{e} \frac{\partial^2 u_i}{\partial x^2} - v_F^2 \nabla . (\nabla n_{se}) + \frac{\hbar^2}{4m_e^2} \nabla . (\nabla^3 n_{se}) + \frac{A_{xc}}{m_e} \nabla . (\nabla n_{se})$$
$$- \frac{\varepsilon}{4m_e} \nabla . (\nabla |\tilde{E}|^2) = 0. \tag{35}$$

The contribution of lattice ions to the dynamics on the slow scale is described by equations (6) and (16) as,

$$\frac{\partial^2 u_i}{\partial t^2} - \left( c_s^2 + \frac{\beta^2}{\rho_i \varepsilon} \right) \frac{\partial^2 u_i}{\partial x^2} = \left( \frac{\beta e}{\rho_i \varepsilon} \right) n_{se}. \tag{36}$$

Eqs. (31), (35) and (36) represent a set of nonlinear evolution equations describing the coupled dynamics of electron waves with lattice ion vibrations in piezoelectric quantum plasma.

In the pursuit of obtaining solitary solutions for the coupled nonlinear evolution equations (31), (35) and (36), we focus on exploring traveling wave solutions, which have been extensively investigated. To facilitate our analysis, we introduce a co-moving frame $\xi = x - v_g t$, moving with velocity, $v_g = \frac{\partial \omega}{\partial k} = \frac{v_F^2 k}{\omega_0}$, where $\xi$ is the new coordinate, $x$ is the original coordinate and $v_g$ represents the group velocity.

Introducing $\tilde{E} \Box e^{i\Delta \omega t}$ in eq. (31) and eliminating $u_i$ from equations (32) & (33), we get

$$2\Delta \omega \tilde{E} + \frac{1}{2} \frac{\partial v_g}{\partial k} \frac{\partial^2 \tilde{E}}{\partial \xi^2} - \frac{n_{se} . \omega_p}{2n_0} \tilde{E} - \frac{\hbar^2}{8\omega_0 m_e^2} \frac{\partial^4 \tilde{E}}{\partial \xi^4} - \frac{A_{xc}}{2\omega_0 m_e} \frac{\partial^2 \tilde{E}}{\partial \xi^2} = 0, \tag{37}$$

and

$$\left[ \left( v_g^2 - v_F^2 - \frac{A_{xc}}{m_e} \right) \frac{\partial^2}{\partial \xi^2} + \frac{\hbar^2}{4m_e^2} \frac{\partial^4}{\partial \xi^4} + \omega_p^2 + \frac{\omega_p^2 \beta^2}{\rho_i \varepsilon (v_g^2 - c_s^2) - \beta^2} \right] n_{se} = \frac{\varepsilon}{4m_e} \frac{\partial^2 |\tilde{E}|^2}{\partial \xi^2}. \tag{38}$$

Eqs. (37) and (38) represent the modified Quantum Zakharov equations governing the coupling between the electron waves and lattice ion vibrations in the presence of piezoelectric field. In the static limit, these equations reduce to the nonlinear Schrödinger equation, which admits soliton solutions. Under this limit for our case, we consider

$$\left(v_g^2 - v_F^2 - \frac{A_{xc}}{m_e}\right)\frac{\partial^2}{\partial \xi^2} + \frac{\hbar^2}{4m_e^2}\frac{\partial^4}{\partial \xi^4} \ll \frac{\omega_p^2 \beta^2}{\rho_i \varepsilon \left(v_g^2 - c_s^2\right) - \beta^2}$$, to obtain the expression for perturbed

number density from eq. (38) as,

$$n_{se} = \frac{\varepsilon}{4m_e \omega_p^2}\left[1 - \frac{\beta^2}{\rho_i \varepsilon \left(v_g^2 - c_s^2\right) - \beta^2}\right]\frac{\partial^2 |\tilde{E}|^2}{\partial \xi^2}. \tag{39}$$

In the system described by the above equation, areas where the electric field has larger amplitudes are associated with regions where the local number density of particles is significantly reduced. On substituting the expression for the perturbed number density from eq. (39) into eq. (37), we get

$$2\Delta\omega \tilde{E} + \frac{v_F^2}{2\omega_0}\frac{\partial^2 \tilde{E}}{\partial \xi^2} - \frac{\hbar^2}{8\omega_0 m_e^2}\frac{\partial^4 \tilde{E}}{\partial \xi^4} - \frac{A_{xc}}{2\omega_0 m_e}\frac{\partial^2 \tilde{E}}{\partial \xi^2}$$
$$- \left(\frac{\omega_p}{2n_0}\right)\left[\frac{\varepsilon}{4m_e \omega_p^2}\left(1 - \frac{\beta^2}{\rho_i \varepsilon \left(v_g^2 - c_s^2\right) - \beta^2}\right)\right]\tilde{E}\frac{\partial^2 |\tilde{E}|^2}{\partial \xi^2} = 0. \tag{40}$$

Integrating the above equation and applying the boundary condition that all the perturbations vanish at infinity. i.e, $\tilde{E} \to 0$ as $\xi \to \infty$, we get

$$\frac{\partial \xi}{\partial \tilde{E}} = \sqrt{\frac{6}{a}(1-b')\left(\frac{1}{\tilde{E}^2} - \frac{b}{3(1-b')}\right)}, \tag{41}$$

where,

$$a = \left(\frac{4\omega_0 \Delta\omega}{v_F^2}\right),$$

$$b = \frac{\varepsilon}{\omega_p m_e v_F^2}\left(\frac{\omega_p}{2n_0}\right)\left(1 - \frac{\beta^2}{\rho_i \varepsilon \left(v_g^2 - c_s^2\right) - \beta^2}\right),$$

$$b' = \frac{A_{xc}}{m_e v_F^2}.$$

The above expression shows that $\partial \tilde{E}/\partial \xi \to \infty$ at the maxima $(\xi = 0)$, thus $\tilde{E} = 1\left/\sqrt{\frac{b}{3(1-b')}}\right.$.

Integrating eq. (41), we get

$$\xi = \sqrt{\frac{6}{a}(1-b')}\left[\sqrt{1-\left(\frac{b}{3(1-b')}\right)\tilde{E}^2} + \log\left|\left(\sqrt{\frac{b}{3(1-b')}}\right)\tilde{E}\right| - \log\left|1+\sqrt{1-\left(\frac{b}{3(1-b')}\right)\tilde{E}^2}\right|\right], \quad (42)$$

which represents a singular spiky soliton solution, also known as Cusp Soliton. This type of soliton is characterized by sharp spikes or cusps in its shape, and it plays a significant role in the dynamics of the coupled electron waves and lattice ion vibrations in the piezoelectric semiconductor quantum plasma [26]. The spiky solitons observed are likely a result of the nonlinearities introduced by quantum effects such as Fermi pressure, Bohm potential, and exchange-correlation forces. These quantum factors create steeper density gradients and sharper potential wells, which lead to solitons with sharp, spiky profiles. The spiky solitons can be physically significant in the sense that it highlights the sensitivity of the system to small perturbations and the strong localization of energy.

Fig. 3 shows the variation of the field amplitude of cusp soliton profile for different values of piezoelectric coupling constant. The piezoelectric coupling constant typically falls within the range of 0.045 $C/m^2$ to 0.35 $C/m^2$ [46]. Here, the solid line shows the variation for $\beta = 0.054$ $C/m^2$ and the dashed line shows the trend for $\beta = 0.21$ $C/m^2$, for InSb and GaAs respectively. A significant increment can be seen in the amplitude of the electric field by decreasing the strength of the piezoelectric coupling constant in piezoelectric semiconductor quantum plasmas.

Fig. 4, shows the variation of the field amplitude of cusp soliton profile for different value of electron density at piezoelectric coupling constant 0.21 $C/m^2$. The solid line shows the variation for $n = 10^{28} m^{-3}$ and the dashed line shows the variation for $n = 10^{27} m^{-3}$. In this case, increment can be seen in the amplitude of the electric field by increasing the number density in piezoelectric semiconductor quantum plasmas. This observation is consistent with eq. (38), which indicates that the perturbed density is proportional to the second derivative of the electric field. As the electron density increases, the number of electrons contributing to the plasma wave also increases leading stronger interactions between lattice vibrations and plasma wave, resulting in higher electric field amplitude.

Fig. 5, shows a comparison between the profiles of cusp soliton profile in presence of quantum effects (solid line) and in absence of quantum effects $(\hbar \to 0)$ (dashed line). It can be seen from the figure that the electric field amplitude of the cusp soliton's profile experiences an increment of 1.71 times as compared to the case where quantum effects are ignored, due to quantum correction terms including electron Fermi pressure, quantum Bohm potential and exchange-correlation potential in the piezoelectric semiconductor quantum plasma. These quantum correction terms lead to redistribution of energy and higher electron occupancy in elevated energy states, resulting in overall amplification of electric field associated with soliton's profile in semiconductor plasma.

## 4. Summary and discussion

We have explored lattice ions vibrations and electron waves coupling in piezoelectric semiconductor quantum plasmas using a two-time scale theory. Piezoelectric effects, nonlinear phenomena, and plasma effects in semiconductors have been extensively investigated due to their broad technological applications. A dispersion relation for coupled wave is derived using QHD fluid model incorporating the quantum effects of electron's Fermi pressure, the quantum Bohm potential by using non-dimensional quantum parameter and the exchange and correlation potential. Further, a set of nonlinear evolution equation has been derived using two-time scale theory and soliton solution for these coupled nonlinear evolution equations has been obtained using modified quantum Zakharov equations and its variation with number density and piezoelectric coupling strength have been analyzed. Also, the variation in profile of cusp solitons in quantum regime have been studied and compared with the case where quantum effects are absent.

It is observed that the transmission in quantum plasmas exhibits a gradual and less abrupt decline, due to the incorporation of quantum effects, in contrast to classical plasmas and the power transmission improves by 13% as quantum parameter increases, driven by the escalation of quantum degeneracy in response to higher plasma density showing the collective effects of Fermi pressure and Bohm potential. On the basis of outcomes of our work, it turns out that the electric field amplitude of cusp soliton's profile experiences a rise of 1.71 times as compared to the case where quantum effects are absent by the inclusion of quantum correction term due to electron's Fermi pressure, quantum Bohm potential and the exchange and correlation potential. These correction terms induce a redistribution of energy and increased electron occupancy in elevated energy states, collectively strengthening the stability and dynamics of solitons leading to amplification of the electric field linked to soliton profiles. It is found that the soliton's field amplitude increases significantly with particle density and by decreasing coupling strength in piezoelectric semiconductor quantum plasma. Quantum effects influences the electron occupancy in energy states and, consequently contribute to soliton stability. This study will be helpfull in designing and optimizing piezoelectric nanoelectronic devices especially for the advancement of energy harvesting and conversion technologies.

**Declaration of Competing Interest**

The authors report no declarations of interest.

**Acknowledgement**

The authors thank SERB – DST, Govt. of India for financial support under MATRICS scheme (grant no. : MTR/2021/000471).

**References**

[1] J. Curie, P. Curie, Développement, par pression, de l'électricité polaire dans les cristaux hémièdres à faces inclines, Comptes Rendus 91 (1880) 294.

[2] G. Lippmann, Relations entre les phénomènes électriques et capillaires, Ann. Chim. Phys. 5 (1875) 494–549.

[3] W.G. Cady, Piezoelectricity: An Introduction to the Theory and Applications of Electromechanical Phenomena in Crystals, McGraw-Hill.

[4] W.P. Mason, Piezoelectricity, its history and applications, J. Acoust. Soc. Am. 70 (1981) 1561–1566.

[5] W. Voigt, On an apparently necessary extension of the theory of elasticity, Ann. Phys. 52 (1894) 536.

[6] A.R. Hutson, Piezoelectricity and conductivity in ZnO and CdS, Phys. Rev. Lett. 4 (1960) 505.

[7] S. Roundy, P.K. Wright, A piezoelectric vibration based generator for wireless electronics, Smart Mater. Struct. 13 (2004) 1131.

[8] E. Lefeuvre, A. Badel, C. Richard, D. Guyomar, Piezoelectric energy harvesting device optimization by synchronous electric charge extraction, J. Intell. Mater. Syst. Struct. 16 (2005) 865–876.

[9] H.A. Sodano, D.J. Inman, G. Park, Comparison of piezoelectric energy harvesting devices for recharging batteries, J. Intell. Mater. Syst. Struct. 16 (2005) 799–807.

[10] K.A. Cook-Chennault, N. Thambi, A.M. Sastry, Powering MEMS portable devices—a review of non-regenerative and regenerative power supply systems with special emphasis on piezoelectric energy harvesting systems, Smart Mater. Struct. 17 (2008) 043001.

[11] Z.L. Wang, X. Wang, J. Song, J. Liu, Y. Gao, Piezoelectric nanogenerators for self-powered nanodevices, IEEE Pervasive Comput. 7 (2008) 49–55.

[12] P. Li, F. Jin, J. Ma, One-dimensional dynamic equations of a piezoelectric semiconductor beam with a rectangular cross section and their application in static and dynamic characteristic analysis, Appl. Math. Mech. 39 (2018) 685–702.

[13] S.R. Anton, H.A. Sodano, A review of power harvesting using piezoelectric materials, Smart Mater. Struct. 16 (2007) R1.

[14] J. Twiefel, H. Westermann, Survey on broadband techniques for vibration energy harvesting, J. Intell. Mater. Syst. Struct. 24 (2013) 1291–1302.

[15] G. Manfredi, P.A. Hervieux, J. Hurst, Fluid descriptions of quantum plasmas, Rev. Mod. Plasma Phys. 5 (2021) 1–38.

[16] G. Manfredi, F. Haas, Self-consistent fluid model for a quantum electron gas, Phys. Rev. B 64 (2001) 075316.

[17] F. Haas, L.G. Garcia, J. Goedert, G. Manfredi, Quantum ion-acoustic waves, Phys. Plasmas 10 (2003) 3858–3866.

[18] P. Kumar, S. Singh, N. Ahmad, Beam-plasma streaming instability in spin-polarized quantum magnetoplasma, Phys. Scr. 95 (2020) 075604.

[19] A. Mehramiz, J. Mahmoodi, S. Sobhanian, Approximation method for a spherical bound system in the quantum plasma, Phys. Plasmas 17 (2010) 082110.

[20] A.P. Misra, Electromagnetic surface modes in a magnetized quantum electron-hole plasma, Phys. Rev. E 83 (2011) 057401.

[21] S. Ghosh, S. Dubey, R. Vanshpal, Quantum effect on parametric amplification characteristics in piezoelectric semiconductors, Phys. Lett. A 375 (2010) 43–47.

[22] I. Zeba, M.E. Yahia, P.K. Shukla, M.W. Moslem, Electron–hole two-stream instability in a quantum semiconductor plasma with exchange-correlation effects, Phys. Lett. A 376 (2012) 2309–2313.

[23] I. Zeba, C. Uzma, M. Jamil, M. Salimullah, P.K. Shukla, Colloidal crystal formation in a semiconductor quantum plasma, Phys. Plasmas 17 (2010) 032105.

[24] M.E. Yahia, I.M. Azzouz, M.W. Moslem, Quantum effects in electron beam pumped GaAs, Appl. Phys. Lett. 103 (2013) 082105.

[25] A. Sharma, N. Yadav, S. Ghosh, Modified acousto-electric interactions in colloids laden semiconductor quantum plasmas, Int. J. Sci. Res. Publ. 3 (2013) 1–7.

[26] S. Ali, M.W. Moslem, P.K. Shukla, R. Schlickeiser, Linear and nonlinear ion-acoustic waves in an unmagnetized electron-positron-ion quantum plasma, Phys. Plasmas 14 (2007) 082307.

[27] M.R. Amin, Modulation of a quantum positron acoustic wave, Astrophys. Space Sci. 359 (2015) 1–9.

[28] C. Uzma, I. Zeba, H.A. Shah, M. Salimullah, Stimulated Brillouin scattering of laser radiation in a piezoelectric semiconductor: Quantum effect, J. Appl. Phys. 105 (2009) 013307.

[29] I. Zeba, C. Uzma, M. Jamil, M. Salimullah, P.K. Shukla, Colloidal crystal formation in a semiconductor quantum plasma, Physics of Plasmas 17 (2010) 032105.

[30] S.V. Vladimirov, Y.O. Tyshetskiy, On description of a collisionless quantum plasma, Phys. Usp. 54 (2011) 1243.

[31] P.K. Shukla, B. Eliasson, Nonlinear aspects of quantum plasma physics, Usp. Fiz. Nauk 53 (2010) 51.

[32] P.K. Shukla, S. Ali, L. Stenflo, M. Marklund, Nonlinear wave interactions in quantum magnetoplasmas, Phys. Plasmas 13 (2006) 112111.

[33] Y. Wang, X. Lü, Modulational instability of electrostatic acoustic waves in an electron-hole semiconductor quantum plasma, Phys. Plasmas 21 (2014) 2.

[34] M. Bonitz, Z.A. Moldabekov, T.S. Ramazanov, Quantum hydrodynamics for plasmas—Quo vadis?, Phys. Plasmas 26 (2019) 090601.

[35] K. Sharma, U. Deka, Comprehensive review on various instabilities in semiconductor quantum plasma, Braz. J. Phys. 51 (2021) 1944–1955.

[36] D.L. White, Amplification of ultrasonic waves in piezoelectric semiconductors, J. Appl. Phys. 33 (1962) 2547.

[37] P. Kumar, N. Ahmad, Surface plasma wave in spin-polarized semiconductor quantum plasma, Laser Part. Beams 38 (2020) 159–164.

[38] J.F. Haskins, J.S. Hickman, A derivation and tabulation of the piezoelectric equations of state, J. Acoust. Soc. Am. 22 (1950) 584–588.

[39] R. Bechmann, The linear piezoelectric equations of state, Br. J. Appl. Phys. 4 (1953) 210.

[40] A.R. Hutson, D.L. White, Elastic wave propagation in piezoelectric semiconductors, J. Appl. Phys. 33 (1962) 40–47.

[41] P.K. Shukla, B. Eliasson, Novel attractive force between ions in quantum plasmas, Phys. Rev. Lett. 108 (2012) 165007.

[42] Z. Moldabekov, T. Schoof, P. Ludwig, M. Bonitz, T. Ramazanov, Statically screened ion potential and Bohm potential in a quantum plasma, Phys. Plasmas 22 (2015) 10.

[43] S.C. Li, The effects of Bohm potential on ion-acoustic solitary waves interaction in a nonplanar quantum plasma, Phys. Plasmas 17 (2010) 8.

[44] A. Rasheed, M. Jamil, M. Mirza, G. Murtaza, Large amplitude electrostatic solitons in spin-polarized electron–positron–ion quantum plasma, Astrophys. Space Sci. 336 (2011) 535–540.

[45] M. Bonitz, A. Filinov, J. Böning, J.W. Dufty, Introduction to Complex Plasmas, Springer, Berlin, Heidelberg (2010).

[46] S. Ghosh, P. Khare, Effect of density gradient on the acousto-electric wave instability in ion-implanted semiconductor plasmas, Acta Phys. Pol. A 109 (2006) 187–197.

[47] B.R. Nag, Theory of Electrical Transport in Semiconductors, Pergamon Press, Oxford (1972).

[48] M. Hassel, H.S. Kwok, Picosecond Phenomena III, Springer Series in Chemical Physics, New York (1982).

[49] R.L. Kallaher, J.J. Heremans, Spin and phase coherence measured by antilocalization in n-InSb thin films, Phys. Rev. B 79 (2009) 075322.

[50] K. Seeger, Semiconductor Physics, Springer-Verlag, New York (1973).

[51] G. Arlt, P. Quadflieg, Piezoelectricity in III–V compounds with a phenomenological analysis of the piezoelectric effect, Phys. Status Solidi (b) 25 (1968) 323–330.

[52] K. Nisihikawa, H. Hojo, K. Mima, H. Ikezi, Coupled nonlinear electron-plasma and ion-acoustic waves, Phys. Rev. Lett. 33 (1974) 148.

[53] R.O. Dendy, Plasma Dynamics, Oxford University Press (1990).

[54] S.G. Thornhill, D. TerHaar, Langmuir turbulence and modulational instability, Phys. Rep. 43 (1978) 3–99.

# Figure captions

Fig. 1  Variation of $\omega/\omega_p$ with $kc/\omega_p$ in piezoelectric semiconductor quantum plasma and in absence of quantum effects $(\hbar = 0)$ for $n_{0e} = 10^{28} m^{-3}$, $\beta = 0.054 C/m^2$.

Fig. 2  Variation of $\omega/\omega_p$ with $kc/\omega_p$ for different value of nondimensional quantum parameter $H$ for $n_{0e} = 10^{28} m^{-3}$, $\beta = 0.054 C/m^2$.

Fig. 3  Profile of cusp soliton with the variation in piezoelectric coupling constant $\beta$ for $n_{0e} = 10^{28} m^{-3}$.

Fig. 4  Profile of cusp soliton with the variation in number density $n_{0e}$ for $\beta = 0.054 C/m^2$.

Fig. 5  comparison of the profiles of cusp solitons in piezoelectric semiconductor quantum plasma and in absence of quantum effects $(\hbar = 0)$ for $n_{0e} = 10^{28} m^{-3}$, $\beta = 0.054 C/m^2$.

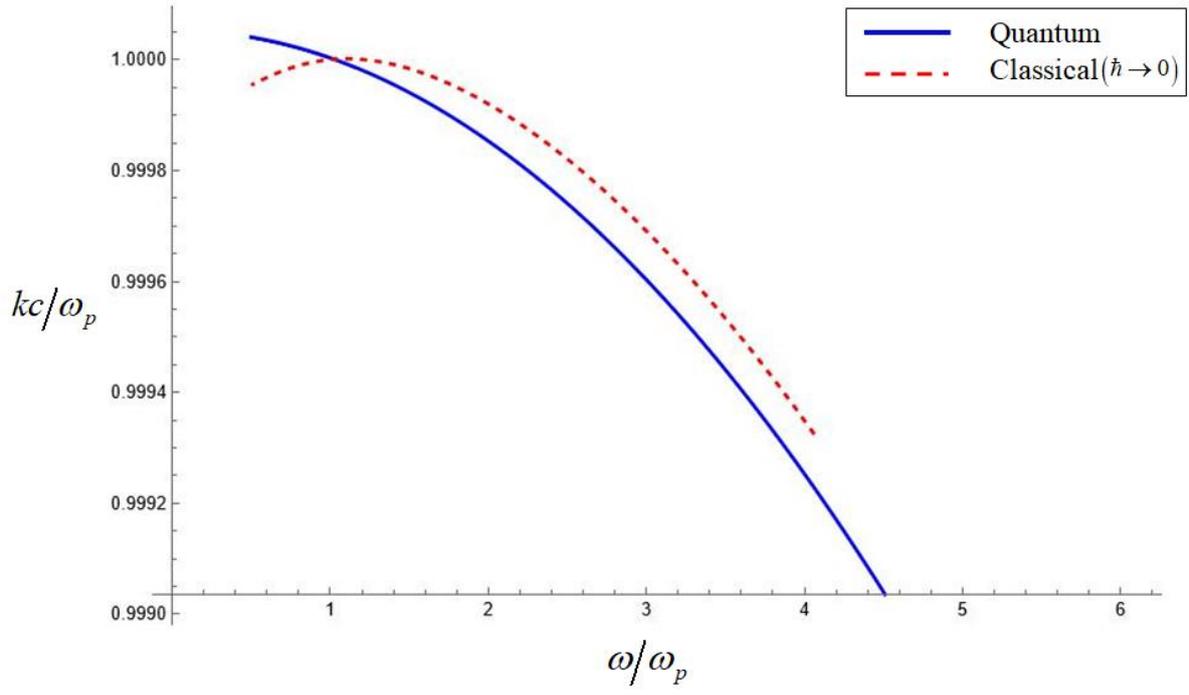

**Fig. 1**

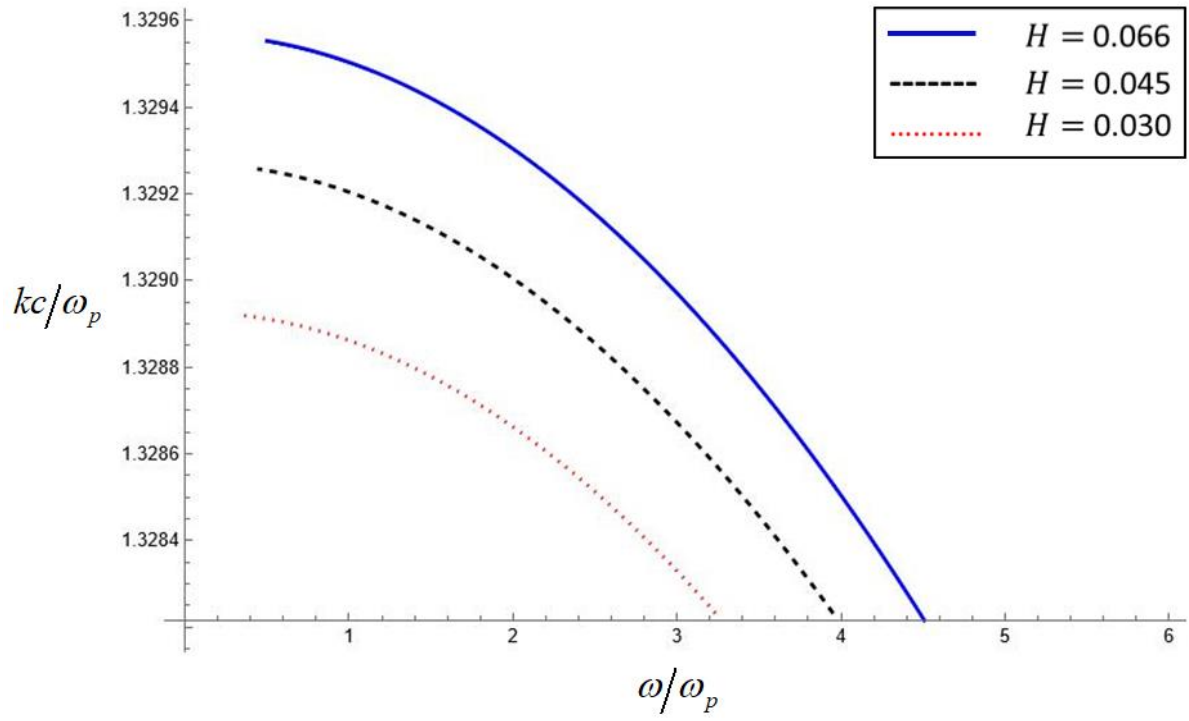

**Fig. 2**

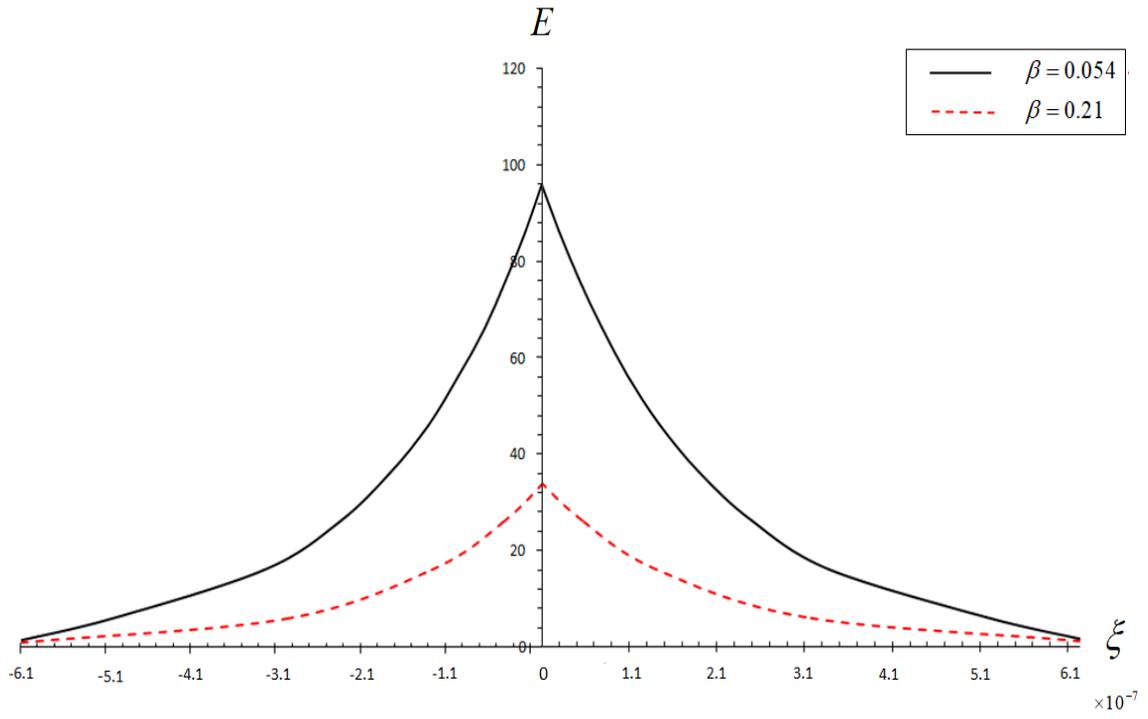

**Fig. 3**

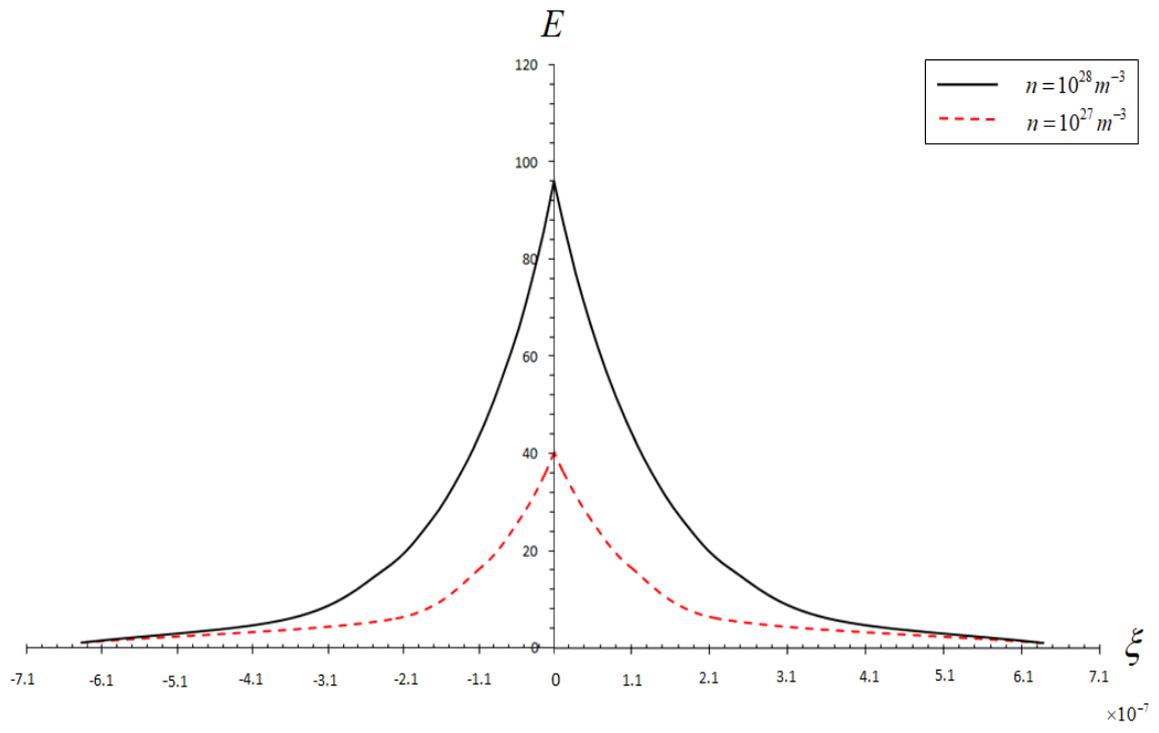

**Fig. 4**

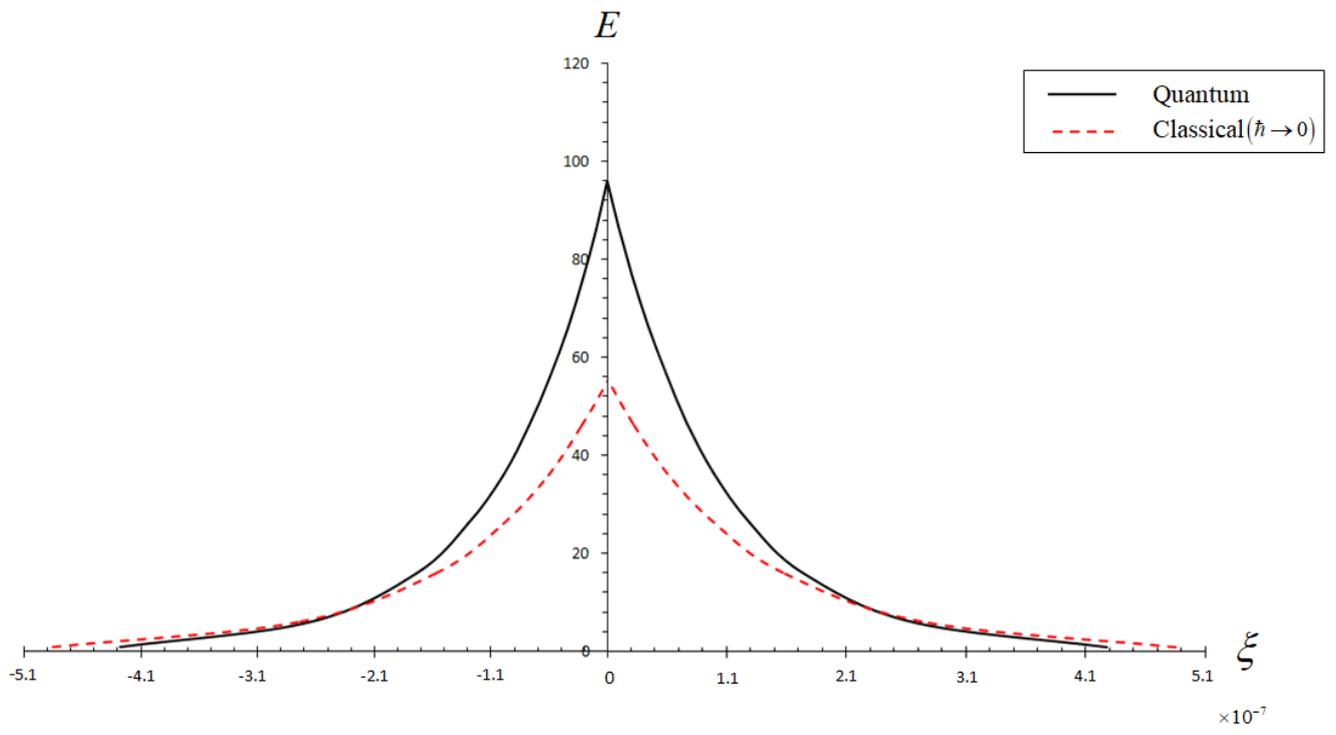

**Fig. 5**